\documentclass[12pt]{article}

\usepackage{epsfig}
\usepackage{amsfonts}
\usepackage{amsopn}
\usepackage{amsmath}
\usepackage{amssymb}

\allowdisplaybreaks

\begin{document}

\baselineskip=18.8pt plus 0.2pt minus 0.1pt

%%%%%%%%%%% Private Macros %%%%%%%%%%%%%
\makeatletter

\@addtoreset{equation}{section}
\renewcommand{\theequation}{\thesection.\arabic{equation}}
\renewcommand{\thefootnote}{\fnsymbol{footnote}}
\newcommand{\beq}{\begin{equation}}
\newcommand{\eeq}{\end{equation}}
\newcommand{\bea}{\begin{eqnarray}}
\newcommand{\eea}{\end{eqnarray}}
\newcommand{\nn}{\nonumber\\}
\newcommand{\hs}[1]{\hspace{#1}}
\newcommand{\vs}[1]{\vspace{#1}}
\newcommand{\p}{\partial}
\newcommand{\bra}[1]{\left\langle  #1 \right\vert }
\newcommand{\ket}[1]{\left\vert #1 \right\rangle }
\newcommand{\vev}[1]{\left\langle  #1 \right\rangle }

\newcommand{\eqnum}{\addtocounter{equation}{1}\tag*{\normalsize{(\thesection.\arabic{equation})}}}

\makeatother
%%%%%%%%% End of private macros %%%%%%%%%%%

%%%%%%%%%%%%%%%%%%%%%%%%%%%%%%%%%%%%%%%%%%%%%%%%%%%%%%%%%%%%%%%%%%%
\begin{titlepage}
\title{
\hfill\parbox{4cm}
{\normalsize\tt AEI-2012-086 
}\\
\vspace{1cm}
Towards the Ground State of the Supermembrane
}
\author{Yoji Michishita
\thanks{
{\tt michishita@edu.kagoshima-u.ac.jp}
}
\\[7pt]
{\it Department of Physics, Faculty of Education, Kagoshima University}\\
{\it Kagoshima, 890-0065, Japan}
\\[7pt]
Maciej Trzetrzelewski
\thanks{
{\tt maciej.trzetrzelewski@gmail.com}
}
\\[7pt]
{\it Max Planck Institute for Gravitational Physics (Albert Einstein Institute)}\\
{\it Am M\"uhlenberg 1, D-14476, Golm, Germany
}
}

\date{\normalsize August, 2012}
\maketitle
\thispagestyle{empty}

\begin{abstract}
\normalsize
The explicit, near the origin, form of the ground state of the $SU(2)$ 
supermembrane matrix model is studied. We evaluate the 2nd order terms of the Taylor expansion of the wave-function, which together with the 0th and the 1st order terms 
can be used to determine  other terms by recurrence equations coming from the Schr\"odinger equation.
\end{abstract}
\end{titlepage}

\clearpage
%%%%%%%%%%%%%%%%%%%%%%%%%%%%%%%%%%%%%%%%%%%%%%%%%%%%%%%%%%%%%%%%%%%
%%%%%%%%%%%%%%%%%%%%%%%%%%%%%%%%%%%%%%%%%%%%%%%%%%%%%%%%%%%%%%%%%%%
\section{Introduction}

After more than two decades, since the formulation of the supermembrane matrix model \cite{model1,model2,model3}, the existence of the zero-energy ground state of the theory, as well as its explicit construction, are still open issues. Any solution to this problem results in long-standing implications - if the normalizable state does not exist the theory is likely to be meaningless. However even in such worst case scenario it is still possible that the large $N$ limit results in the function that \emph{is} still normalizable \cite{spikes}. Therefore the supermembrane could make sense even though its regularization is ill defined.
Clearly, the existence of the ground state of the model has important consequences not only for membranes but also for string theory due to the BFSS conjecture \cite{bfss}.

Although there are strong indications, based on the Witten index calculations \cite{index1,index2}, that the ground state exists, the fact that the spectrum of the model is continuous \cite{dWLN,smilga} makes the index ill-defined and hence cannot serve directly as a rigorous proof of the existence of the state (for a more detailed discussion see e.g. \cite{douglas}). 

There are however other techniques, not relying on the supersymmetric index, which make the proof accessible. A notable example of this kind is the deformation technique \cite{porrati,solovej} which was used in a different but related matrix model (corresponding to D0-D4 bound states). Other promising approach is based on the group averaging techniques - in particular in references \cite{groupav1,groupav2} it was shown that the question about the existence of the ground state can be answered using a simpler model with two interacting matrices (while in the original model there are  nine of them). Such tremendous simplification was possible due to the hidden octonionic structure of the model.

In this paper, rather than focusing on the existence, we address the question about the explicit form of the ground state. Although its asymptotic form is very well studied \cite{asym1,asym2} the corresponding behavior near the origin is still not known to a satisfactory degree. Performing the Taylor expansion of the ground state about $X=0$, the 0th order term (i.e. the coordinate independent one) for the $SU(2)$ model has been constructed explicitly \cite{construction1} and proven to be unique \cite{construction2,construction3} which confirmed earlier symbolic results using Mathematica \cite{wosiek}. The 1st order term is now also available and turns out to be unique as well \cite{construction4}. Because the zero-energy state $\ket{\psi}$ satisfies (schematically) $(\partial_X + X^2)\ket{\psi}=0$, the 0th, the 1st \emph{and} the 2nd order terms  are crucial in finding the higher order terms 
by an appropriate recurrence equation. It is therefore important to find the remaining 2nd order term.
After summarizing notation and basic facts in section 2 and 3, we shall determine that term explicitly in section 4.
We find that there are two independent terms of this sort. Since explicit expressions of those states are lengthy,
we give them in the Appendix.

%%%%%%%%%%%%%%%%%%%%%%%%%%%%%%%%%%%%%%%%%%%%%%%%%%%%%%%%%%%%%%%%%%%
%%%%%%%%%%%%%%%%%%%%%%%%%%%%%%%%%%%%%%%%%%%%%%%%%%%%%%%%%%%%%%%%%%%
\section{Preliminaries}

The supermembrane matrix model is a quantum mechanical system with $\mathcal{N} = 16$ supersymmetries, $SU(N)$ gauge invariance (in this paper we consider $N=2$) and $Spin(9)$ symmetry. The theory involves real bosonic variables $X_i^a$ (the coordinates) and real fermionic ones $\theta_{\alpha}^a$ (Majorana spinors) with $i = 1,\ldots, 9$, $\alpha = 1,\ldots,16$ and
$a = 1, \ldots, N^2-1$ - spatial, spinor and color indices respectively. The corresponding supercharges and the Hamiltonian of the model are
\begin{equation} \label{Q}
Q_{\alpha}= {\rm Tr}\left(P_i \gamma^i \theta + \frac{i}{2}[X_i,X_j]\gamma^{ij}\theta\right)_{\alpha}, \ \ \ \ \ \{Q_{\alpha},Q_{\beta}\}=2\delta_{\alpha\beta}H,
\end{equation}
where $\gamma^i$ are $16 \times 16$ real, gamma matrices such that $\{\gamma^i,\gamma^j\}=2\delta^{ij}\bold{1}$ and $\gamma^{ij}=\frac{1}{2}[\gamma^i,\gamma^j]$. 
The Hilbert space consists of all the states $\ket{s}$ satisfying the singlet constraint 
\beq
G_a\ket{s}=0, \ \ \  G_a=f_{abc}(X_i^b P_i^c+i\theta_\alpha^b\theta_\alpha^c),
\eeq
where $P_i^a$ denote the conjugate momenta i.e. $[X_i^a,P_j^b]=i\delta_{ij}\delta^{ab}$, and $f_{abc}$ are the structure constants of $SU(N)$. The trace in (\ref{Q}) is over the $SU(N)$ matrix given by $X_i=X_i^aT_a$,  $P_i=P_i^aT_a$ and $\theta_\alpha = \theta_{\alpha}^aT_a$ where $T_a$'s are the basis elements of the group algebra. For more details of the model we refer to existing reviews in the literature \cite{rev1,rev2}.

Let $\ket{\psi}$ denote the conjectured ground state i.e. a normalizable vector s.t. $Q_{\alpha}\ket{\psi}=0$. 
It has been shown that $\ket{\psi}$ must be $SO(9)$ singlet \cite{hh02}.
When we expand $\ket{\psi}$ in the coordinates $X^a_i$:
\bea
\ket{\psi} & = & \ket{\phi} + X^a_i \ket{\phi^a_i} + X^{a_1}_{i_1}X^{a_2}_{i_2}\ket{\phi^{a_1 a_2}_{i_1i_2}} + \ldots
\nn
& = & \sum_{n=0}^{\infty}X_{i_1}^{a_1}\ldots X_{i_n}^{a_n}\ket{\phi_{i_1\ldots i_n}^{a_1 \ldots a_n}},
\eea
coordinate independent states $\ket{\phi^{a_1\ldots a_n}_{i_1\ldots i_n}}$, which are constructed by acting creation
operators made of $\theta^a_{\alpha}$ on the vacuum for those operators, play an important role. In our case of $SU(2)$
gauge group, classification of the coordinate independent states by representations has been given in \cite{construction3}. 

Zero-energy state equation $Q_{\alpha}\ket{\psi}=0$ can be decomposed into three independent sequences $m = 0, 1, 2$ relating $\ket{\phi^{a_1\ldots a_{3n+m}}_{i_1\ldots i_{3n+m}}}$ with $\ket{\phi^{a_1\ldots a_{3(n+1)+m}}_{i_1\ldots i_{3(n+1)+m}}}$, while the first two of those equations contain only one $\ket{\phi^{a_1\ldots a_n}_{i_1\ldots i_n}}$:
\begin{equation} \label{1st}
\gamma^i \theta^a \ket{\phi^a_i} = 0, 
\end{equation}
\begin{equation} \label{2nd}
\gamma^{i_1}\theta^{a_1}\ket{\phi^{a_1a_2}_{i_1i_2}} = 0, 
\end{equation}
\[
[X_i,X_j]^a X^{a_1}_{i_1}\ldots X^{a_{3n+m}}_{i_{3n+m}}\gamma^{ij}\theta^a\ket{\phi^{a_1\ldots a_{3n+m}}_{i_1 \ldots i_{3n+m}}}
\]
\begin{equation} \label{rest}
 =  2(3(n+1)+m)X^{a_2}_{i_2}\ldots X^{a_{3(n+1)+m}}_{i_{3(n+1)+m}}\gamma^{i_1}\theta^{a_1}\ket{\phi^{a_1\ldots a_{3(n+1)+m}}_{i_1 \ldots i_{3(n+1)+m}}},
\end{equation}
where $n = 0, 1, 2,\ldots$ . The first three states $\ket{\phi}$, $\ket{\phi^a_i}$ and $\ket{\phi_{i_1i_2}^{a_1a_2}}$ give the starting points for solving each sequence of equations order by order. 

%%%%%%%%%%%%%%%%%%%%%%%%%%%%%%%%%%%%%%%%%%%%%%%%%%%%%%%%%%%%%%%%%%%
%%%%%%%%%%%%%%%%%%%%%%%%%%%%%%%%%%%%%%%%%%%%%%%%%%%%%%%%%%%%%%%%%%%
\section{0th and 1st Order Terms}

The unique candidate for $\ket{\phi}$ which we denote here by $\ket{S}$, has been constructed in \cite{construction1,wosiek}, and the unique candidate for $\ket{\phi^a_i}$ which satisfies (\ref{1st}) has also been constructed in \cite{construction4}. Thus we have the starting points for the sequences $m = 0$ and $1$. It turns out that the explicit expressions for these states are relatively simple if one works with states corresponding to irreducible representations of $SO(9)$ of dimensions $\bold{44}$(symmetric-traceless representation), $\bold{84}$(3-rank antisymmetric representation) and $\bold{128}$(vector-spinor representation) which we denote here by $\ket{ij}_a$, $\ket{ijk}_a$ and $\ket{\alpha i}_a$ respectively\footnote{Note the subscript $a$ corresponding to the color index - a generic state for $SU(2)$ will have the form $\ket{A}_1\ket{B}_2\ket{C}_3$.}. The state
$\ket{\alpha i}_a$ is Grassmann odd, and satisfies the Rarita-Schwinger constraint $(\gamma^i)_{\beta\alpha}\ket{\alpha i}_a=0$.
Actions of $\theta^a$ on these states are given by
\bea
\theta^a_\alpha\ket{ij}_b & = & -\frac{1}{3}
\big[(\gamma^i)_{\alpha\beta}\ket{\beta j}_a+(\gamma^j)_{\alpha\beta}\ket{\beta i}_a\big]\delta^{ab},
\label{itwr1} \\
\theta^a_\alpha\ket{ijk}_b & = & \frac{1}{\sqrt{3}}(\gamma^{[ij})_{\alpha\beta}\ket{\beta k]}_a\delta^{ab},
\label{itwr2} \\
\theta^a_\alpha\ket{\beta i}_b & = & \Big[-\frac{3}{4}(\gamma^j)_{\alpha\beta}\ket{ij}_a
-\frac{\sqrt{3}}{24}(\gamma^{jkl}\gamma^i-9\delta^{ij}\gamma^{kl})_{\alpha\beta}\ket{jkl}_a\Big]\delta^{ab},
\label{itwr3} 
\eea
where $[ijk]$ denotes antisymmetrization of indices with the factor $1/3!$\footnote{For the definition of states $\ket{ij}_a$, $\ket{ijk}_a$ and $\ket{\alpha i}_a$ we use conventions of \cite{construction4} which differ from the conventions of \cite{construction1}  by normalization factors.}.
For the 0th order term one finds that \cite{construction1}
\beq
\ket{\phi} = \alpha_0 \ket{S},
\eeq
\beq
\ket{S} := -\frac{6}{13}\mathop{|||}_{44} 1 \rangle +\mathop{|||}_{844} 1 \rangle,
\eeq
\bea
\mathop{|||}_{44}1\rangle & := & |ik \rangle_1 |jk \rangle_2 |ij \rangle_3,\\
\mathop{|||}_{844}1\rangle & := & |ijk \rangle_1 |ljk \rangle_2 |il \rangle_3 + |ljk \rangle_1 |il \rangle_2 |ijk \rangle_3 + |il \rangle_1 |ijk \rangle_2 |ljk \rangle_3.
\eea
The overall factor $\alpha_0$ cannot be determined by the condition $Q_\alpha\ket{\psi}=0$ - the only remaining constraint is the norm $\vev{\psi|\psi}=1$ which should be used to fix $\alpha_0$. For the 1st order term a similar expression is more complicated however as it turns out it can be written in an elegant form when using $\ket{S}$. One finds that \cite{construction4}
\beq
\ket{\phi_a^i} = \alpha_1 f_{abc}\theta_b \gamma^i \theta_c \ket{S},
\eeq
where $\alpha_1$ is determined again by $\vev{\psi|\psi}=1$. 

The above result suggests a possibility that all the states $\ket{\phi^{a_1\ldots a_n}_{i_1\ldots i_n}}$ of the Taylor expansion of $\ket{\psi}$ can be obtained as fairly simple expressions containing fermionic operators $\theta_{\alpha}^a$, contracted with gamma matrices and $SU(2)$ invariant tensors,  acting on $\ket{S}$. This assertion, if true, implies that  there exist a gauge invariant and $SO(9)$ invariant function $f(X,\theta)$ 
such that the ground state of the supermembrane can be written as
\begin{equation} \label{ansatz}
\ket{\psi} = f(X,\theta)\ket{S}.
\end{equation}
In the following section we shall confirm that conjecture giving an explicit expression for the second order terms.

%%%%%%%%%%%%%%%%%%%%%%%%%%%%%%%%%%%%%%%%%%%%%%%%%%%%%%%%%%%%%%%%%%%
%%%%%%%%%%%%%%%%%%%%%%%%%%%%%%%%%%%%%%%%%%%%%%%%%%%%%%%%%%%%%%%%%%%
\section{Construction of 2nd Order Terms}

To give the starting point for $m = 2$ satisfying (\ref{2nd}), we construct all the candidates for $\ket{\phi^{ab}_{ij}}$ first. For the construction the table of representations in the coordinate independent state space (Table 1) given in \cite{construction3} is quite useful. From the table we will see that there are five candidate representations 
\[
\delta^{ab}\delta^{ij}\ket{S},  \ \ \ \delta^{ab}\ket{\phi^{ij}},  \ \ \ \ket{\phi_{(1)ab}^{ij}},  \ \ \ 
 \ket{\phi_{(2)ab}^{ij}}, \ \ \ \ket{\phi_{(3)ab}^{ij}}.
 \]
Moreover the table tells us that there are two independent solutions to (\ref{2nd}), and indeed we will find that two linear combinations of the above five representations satisfy (\ref{2nd}).  We now explain the details of these states and solutions:
$\ket{\phi_{ab}^{ij}}$, which satisfies $\ket{\phi_{ab}^{ij}}=\ket{\phi_{ba}^{ji}}$,
can be decomposed into the following five irreducible representations of $SU(2)\times SO(9)$:
\bea
(t, t) & : & \ket{\phi_{cc}^{kk}} \nn
(t, s) & : & \ket{\phi_{cc}^{(ij)}}-\frac{1}{9}\delta^{ij}\ket{\phi_{cc}^{kk}} \nn
(s, t) & : & \ket{\phi_{(ab)}^{kk}}-\frac{1}{3}\delta_{ab}\ket{\phi_{cc}^{kk}} \nn
(s, s) & : & \ket{\phi_{(ab)}^{ij}}-\frac{1}{9}\delta^{ij}\ket{\phi_{(ab)}^{kk}}
 -\frac{1}{3}\delta_{ab}\ket{\phi_{cc}^{(ij)}}+\frac{1}{27}\delta_{ab}\delta^{ij}\ket{\phi_{cc}^{kk}} \nn
(a,a) & : & \ket{\phi_{[ab]}^{[ij]}},\nonumber
\eea
where $t, s$, and $a$ mean trace (i.e. singlet), symmetric-traceless, and antisymmetric respectively.
For example, $(t, s)$ stands for ($SU(2)$ singlet)$\times$($SO(9)$ symmetic-traceless) representation.
$(ij)$ and $(ab)$ are symmetrization of indices, and $[ij]$ and $[ab]$ are antisymmetization.

For $SU(2)$, $t, a$, and $s$ correspond to spin 0, 1, and 2 representations respectively.
For $SO(9)$, $t, a$, and $s$ correspond to Dynkin labels [0000], [0100], and [2000]
in Table 1 in \cite{construction3} respectively.
Then the table tells us that in the coordinate independent state space,
\begin{description}
\item[(i)] there is only one $(t, t)$ representation (and therefore proportional to $\ket{S}$).
\item[(ii)] there is only one $(t, s)$ representation.
\item[(iii)] there is no $(s, t)$ representation.
\item[(iv)] there are two $(s, s)$ representations.
\item[(v)] there is only one $(a, a)$ representation.
\end{description}
We have to construct all of these representations explicitly to determine $\ket{\phi_{ab}^{ij}}$.

%%%%%%%%%%%%%%%%%%%%%%%%%%%%%%%%%%%%%%%%%%%%%%%%%%%%%%%%%%%%%%%%%%%
\subsection{Construction of $(t,t), (t,s), (a,a)$, and  $(s,s)$ Representations}

Candidates for most of the above states are given by appropriately symmetrizing, antisymmetrizing, or contracting indices in 
$O_a^iO_b^j\ket{S}$ and subtracting trace part, where $O_a^i:=\epsilon_{abc}\theta^b\gamma^i\theta^c$:
\bea
(t, t) & : & \delta_{ab}\delta^{ij}O_c^kO_c^k\ket{S} \nn
(t, s) & : & \delta_{ab}\ket{\phi^{ij}}:=
 \delta_{ab}\big[O_c^{(i}O_c^{j)}-\frac{1}{9}\delta^{ij}O_c^kO_c^k\big]\ket{S} \nn
(s, t) & : & \delta^{ij}\big[O_{(a}^kO_{b)}^k-\frac{1}{3}\delta_{ab}O_c^kO_c^k\big]\ket{S} \nn
(s, s) & : & \ket{\phi_{(2)}{}^{ij}_{ab}}:=\big[O_{(a}^{(i}O_{b)}^{j)}-\frac{1}{9}\delta^{ij}O_{(a}^kO_{b)}^k
 -\frac{1}{3}\delta_{ab}O_c^{(i}O_c^{j)}+\frac{1}{27}\delta_{ab}\delta^{ij}O_c^kO_c^k\big]\ket{S} \nn
(a, a) & : & \ket{\phi_{(1)}{}^{ij}_{ab}}:= O_{[a}^{[i}O_{b]}^{j]}\ket{S}.\nonumber
\eea
By straightforward calculation using \eqref{itwr1}-\eqref{itwr3} the followings can be shown:
\bea
O_a^iO_a^i\ket{S}=-1440\ket{S}, \\
{}\big[O_{(a}^kO_{b)}^k-\frac{1}{3}\delta_{ab}O_c^kO_c^k]\ket{S}=0,
\eea
as is indicated by the table in \cite{construction3}. Therefore $\ket{\phi^{ij}}$ and $\ket{\phi_{(2)}{}^{ij}_{ab}}$
can be simplified:
\bea
\ket{\phi^{ij}} & = &
 \big[O_c^{(i}O_c^{j)}+160\delta^{ij}\big]\ket{S}, \\
\ket{\phi_{(2)}{}^{ij}_{ab}} & = & \big[O_{(a}^{(i}O_{b)}^{j)}
 -\frac{1}{3}\delta_{ab}O_c^{(i}O_c^{j)}\big]\ket{S}.
\eea
There should be another $(s, s)$ representation, and it will be denoted by $\ket{\phi_{(3)}{}^{ij}_{ab}}$.
Before constructing $\ket{\phi_{(3)}{}^{ij}_{ab}}$, we give explicit expressions of
$\ket{\phi^{ij}}, \ket{\phi_{(1)}{}_{ab}^{ij}}$, and $\ket{\phi_{(2)}{}_{ab}^{ij}}$.
This needs tedious calculation, and we have done it by using Mathematica and
the package for $\gamma$-matrix algebra GAMMA \cite{gran01}.

First, the explicit expression of $O_{(a}^{(i}O_{b)}^{j)}\ket{S}$ is given by
the following:

{ \tiny
\[
O_1^{(i}O_1^{j)}\ket{S} =  \delta^{ij}\Big[
\frac{896}{39}\ket{kl}_1\ket{lm}_2\ket{mk}_3
-\frac{448}{13}\ket{kl}_1\ket{kmn}_2\ket{lmn}_3
-80\ket{kmn}_1\ket{kl}_2\ket{lmn}_3
-80\ket{kmn}_1\ket{lmn}_2\ket{kl}_3
\Big] 
\]
\[
-\frac{1}{39}\Big[
1184\ket{ij}_1\ket{kl}_2\ket{kl}_3
+896\ket{kl}_1\ket{ij}_2\ket{kl}_3
+896\ket{kl}_1\ket{kl}_2\ket{ij}_3
\Big] 
+\frac{160}{3}\Big[
\ket{k(i}_1\ket{j)l}_2\ket{kl}_3
+\ket{k(i}_1\ket{kl}_2\ket{j)l}_3
\Big]
\]
\[
-\frac{3584}{39}\ket{kl}_1\ket{k(i}_2\ket{j)l}_3
+\frac{384}{13}\ket{ij}_1\ket{klm}_2\ket{klm}_3
-\frac{1104}{13}\Big[
\ket{k(i}_1\ket{j)lm}_2\ket{klm}_3
+\ket{k(i}_1\ket{klm}_2\ket{j)lm}_3
\Big] 
\]
\[
+16\Big[
\ket{kl(i}_1\ket{j)m}_2\ket{klm}_3
+\ket{kl(i}_1\ket{klm}_2\ket{j)m}_3
\Big] 
+128\Big[
\ket{klm}_1\ket{m(i}_2\ket{j)kl}_3
+\ket{klm}_1\ket{lm(i}_2\ket{j)k}_3
\Big] 
\]
\beq
+96\Big[
\ket{km(i}_1\ket{kl}_2\ket{lmj)}_3
+\ket{km(i}_1\ket{j)lm}_2\ket{kl}_3
\Big] 
-\frac{16\sqrt{3}}{9}\epsilon^{k_1\dots k_8(i}\Big[\ket{k_1k_2k_3}_1\ket{k_4k_5j)}_2\ket{k_6k_7k_8}_3
+\ket{k_1k_2k_3}_1\ket{k_4k_5k_6}_2\ket{k_7k_8j)}_3
\Big], \eqnum
\eeq

\[
O_{(1}^{(i}O_{2)}^{j)}\ket{S} =
-\frac{1}{351}\Big[
288(\gamma^{k(i})_{\alpha\beta}\big(
\ket{\alpha j)}_1\ket{\beta l}_2\ket{kl}_3+\ket{\alpha l}_1\ket{\beta j)}_2\ket{kl}_3
\big) 
\]
\[
+4448(\gamma^{k(i})_{\alpha\beta}\ket{\alpha l}_1\ket{\beta l}_2\ket{kj)}_3 
+10112\big(
\ket{\alpha (i}_1\ket{\alpha k}_2\ket{kj)}_3-\ket{\alpha k}_1\ket{\alpha (i}_2\ket{j)k}_3
\big)
\Big] 
\]
\[
+\frac{\sqrt{3}}{351}\Big[
904(\gamma^{kl(i})_{\alpha\beta}\big(
\ket{\alpha j)}_1\ket{\beta m}_2\ket{klm}_3+\ket{\alpha m}_1\ket{\beta j)}_2\ket{klm}_3
\big) 
-5552(\gamma^{k})_{\alpha\beta}\big(
\ket{\alpha (i}_1\ket{\beta l}_2\ket{klj)}_3-\ket{\alpha l}_1\ket{\beta (i}_2\ket{klj)}_3
\big) 
\]
\[
-1520\delta^{ij}(\gamma^{k})_{\alpha\beta}\ket{\alpha l}_1\ket{\beta m}_2\ket{klm}_3 
-120\delta^{ij}(\gamma^{klm})_{\alpha\beta}\ket{\alpha n}_1\ket{\beta n}_2\ket{klm}_3 
\]
\beq
-656(\gamma^{klm})_{\alpha\beta}\ket{\alpha (i}_1\ket{\beta j)}_2\ket{klm}_3 
-1040(\gamma^{(i})_{\alpha\beta}\ket{\alpha k}_1\ket{\beta l}_2\ket{klj)}_3 
+1736(\gamma^{kl(i})_{\alpha\beta}\ket{\alpha n}_1\ket{\beta n}_2\ket{klj)}_3
\Big],\eqnum
\eeq
}
\bea
O_2^{(i}O_2^{j)}\ket{S}=O_1^{(i}O_1^{j)}\ket{S}\Big|_{
\ket{*_1}_1\ket{*_2}_2\ket{*_3}_3\rightarrow \ket{*_1}_2\ket{*_2}_3\ket{*_3}_1},
\\
O_3^{(i}O_3^{j)}\ket{S}=O_1^{(i}O_1^{j)}\ket{S}\Big|_{
\ket{*_1}_1\ket{*_2}_2\ket{*_3}_3\rightarrow \ket{*_1}_3\ket{*_2}_1\ket{*_3}_2},
\\
O_{(2}^{(i}O_{3)}^{j)}\ket{S}=O_{(1}^{(i}O_{2)}^{j)}\ket{S}\Big|_{
\ket{*_1}_1\ket{*_2}_2\ket{*_3}_3\rightarrow \ket{*_1}_2\ket{*_2}_3\ket{*_3}_1},
\\
O_{(3}^{(i}O_{1)}^{j)}\ket{S}=O_{(1}^{(i}O_{2)}^{j)}\ket{S}\Big|_{
\ket{*_1}_1\ket{*_2}_2\ket{*_3}_3\rightarrow \ket{*_1}_3\ket{*_2}_1\ket{*_3}_2}.
\eea
In the above, $(i\dots j)$ just means symmetrization of only $i$ and $j$, and does not symmetrize indices between $i$ and $j$. For example,
\beq
(\gamma^{kl(i})_{\alpha\beta}\ket{\alpha n}_1\ket{\beta n}_2\ket{klj)}_3
=\frac{1}{2}\left[(\gamma^{kli})_{\alpha\beta}\ket{\alpha n}_1\ket{\beta n}_2\ket{klj}_3+(\gamma^{klj})_{\alpha\beta}\ket{\alpha n}_1\ket{\beta n}_2\ket{kli}_3\right].
\eeq
We need the following identity to obtain the above expression of $O_1^{(i}O_1^{j)}\ket{S}$:
\bea
0 & = & \epsilon^{k_1\dots k_8i}
\big(\ket{k_1k_2j}_1\ket{k_3k_4k_5}_2\ket{k_6k_7k_8}_3-\ket{k_1k_2k_3}_1\ket{k_4k_5j}_2\ket{k_6k_7k_8}_3
 \nn & &
 +\ket{k_1k_2k_3}_1\ket{k_4k_5k_6}_2\ket{k_7k_8j}_3\big)
 \nn & &
-\frac{1}{3}\delta^{ij}\epsilon^{k_1\dots k_9}\ket{k_1k_2k_3}_1\ket{k_4k_5k_6}_2\ket{k_7k_8k_9}_3,
\eea
which can be shown by using 
$\ket{jk_1k_2}_1=\frac{1}{3!\cdot 6!}\epsilon^{jk_1k_2 l_4\dots l_9}
\epsilon^{l_1\dots l_9}\ket{l_1l_2l_3}_1$.

From the above expression of $O_{(a}^{(i}O_{b)}^{j)}\ket{S}$,
we obtain $\ket{\phi^{ij}}$ and $\ket{\phi_{(2)}{}^{ij}_{ab}}$.
Explicit expression of $\ket{\phi_{(1)}{}^{ij}_{ab}}$ is also obtained by straightforward calculation,
and all of those explicit expressions are summarized in the Appendix.

%%%%%%%%%%%%%%%%%%%%%%%%%%%%%%%%%%%%%%%%%%%%%%%%%%%%%%%%%%%%%%%%%%%
\subsection{Another $(s, s)$ Representation}

Now we have $(t, t), (t, s), (a, a)$, and one of $(s, s)$ representations explicitly.
Then the only missing one is the other $(s ,s)$ state $\ket{\phi_{(3)}{}^{ij}_{ab}}$.
Let us try to construct this representation as states made by acting $\theta$'s on $\ket{S}$,
although it is not clear at present if every state can be constructed in this way.
First, let us consider classifying this kind of states with two symmetrized $SU(2)$ adjoint indices
by the number of $\theta$'s on $\ket{S}$.
In the case of two $\theta$'s, $\theta^a\gamma^{ij}\theta^b\ket{S}$ and $\theta^a\gamma^{ijk}\theta^b\ket{S}$ are
possible. However it is impossible to give two symmetrized $SO(9)$ vector indices to these states.
This is the reason why we did not start with states with two $\theta$'s in the previous subsection.
In the case of four $\theta$'s, two of four adjoint indices of $\theta$'s are contracted, and by Fierz transformation
those contracted indices can be put into the same fermion bilinear. The Fierz transformation may give additional
terms which come from the anticommutation relation of $\theta$'s and have two $\theta$'s. We concentrate on terms with
four $\theta$'s. Then the possible states are
\bea
& & [\theta^a\gamma^{ij}\theta^b][\theta^c\gamma^{kl}\theta^c]\ket{S},\quad
[\theta^a\gamma^{ijk}\theta^b][\theta^c\gamma^{lm}\theta^c]\ket{S},
\nn
& & [\theta^a\gamma^{ij}\theta^b][\theta^c\gamma^{klm}\theta^c]\ket{S},\quad
[\theta^a\gamma^{ijk}\theta^b][\theta^c\gamma^{lmn}\theta^c]\ket{S}.
\eea
Note that $\theta^c\gamma^{kl}\theta^c$ is an $SO(9)$ generator, which annihilates $\ket{S}$.
So the first two states vanish.
If we consider states with two symmetrized $SO(9)$ vector indices,
$[\theta^a\gamma^{kl(i}\theta^b][\theta^c\gamma^{j)kl}\theta^c]\ket{S}$ is the only possibility.
So $\ket{\phi_{(2)}{}^{ij}_{ab}}$ must be proportional to this state (plus terms with less $\theta$'s
and terms for subtracting the trace part),
and indeed is proportional as can be seen from
\bea
\ket{\phi_{(2)}{}^{ij}_{ab}}
& = & -4[\theta^a\gamma^{(i}\theta^c][\theta^b\gamma^{j)}\theta^c]\ket{S}+\text{(terms with two $\theta$'s)}
\nn & & +\text{(terms proportional to $\delta^{ab}$)},
\eea
and the following Fierz transformation:
\bea
{}[\theta^a\gamma^{(i}\theta^c][\theta^b\gamma^{j)}\theta^c]\ket{S} & = & 
-\frac{1}{16\cdot 2}[\theta^a\gamma^{(i}\gamma^{m_1m_2}\gamma^{j)}\theta^b][\theta^c\gamma_{m_1m_2}\theta^c]\ket{S}
\nn & &
-\frac{1}{16\cdot 6}[\theta^a\gamma^{(i}\gamma^{m_1m_2m_3}\gamma^{j)}\theta^b][\theta^c\gamma_{m_1m_2m_3}\theta^c]\ket{S}
\nn & & +\text{(terms with two $\theta$'s)}
\nn & = & 
-\frac{1}{16}[\theta^a\gamma^{kl(i}\theta^b][\theta^c\gamma^{j)kl}\theta^c]\ket{S}
\nn & & +\text{(terms with two $\theta$'s)}+\text{(terms proportional to $\delta^{ij}$)}.
\eea
This shows that using four $\theta$'s we can construct no more $(s,s)$ representation.
Next we consider six $\theta$ case. By Fierz transformation $SU(2)$ adjoint indices are arranged so
that we have two bilinears with $SU(2)$ indices contracted within each of them,
and one bilinear with two symmetrized free indices. For example,
\beq
\ket{\phi'_{(3)}{}^{ij}_{ab}}:=
[\theta^a\gamma^{(i}{}_{n_1}\theta^b][\theta^c\gamma^{j)}{}_{n_2n_3}\theta^c]
[\theta^d\gamma^{n_1n_2n_3}\theta^d]\ket{S}
\eeq
is the only state with two symmetrized $SO(9)$ vector indices, and its traceless part
$\ket{\phi_{(3)}{}^{ij}_{ab}}$ may give another $(s,s)$ representation.
Explicitly,

{\tiny

\[
\ket{\phi'_{(3)}{}^{ij}_{11}}  =
\frac{6144}{13}\delta^{ij}\ket{kl}_1\ket{lm}_2\ket{mk}_3
+21504 \ket{ij}_1\ket{kl}_2\ket{kl}_3
+\frac{3072}{13}\big(
\ket{kl}_1\ket{ij}_2 \ket{kl}_3+\ket{kl}_1\ket{kl}_2\ket{ij}_3
\big) 
\]
\[
+\frac{16384}{13}\big(\ket{ki}_1\ket{lj}_2\ket{kl}_3+\ket{kj}_1\ket{li}_2\ket{kl}_3
+\ket{ki}_1\ket{kl}_2\ket{lj}_3+\ket{kj}_1\ket{kl}_2\ket{li}_3\big)
-\frac{60416}{13}\big(\ket{kl}_1\ket{ki}_2\ket{lj}_3+\ket{kl}_1\ket{kj}_2\ket{li}_3\big)
\]
\[
+\frac{52224}{13}\delta^{ij}\big(
\ket{kmn}_1\ket{lmn}_2\ket{kl}_3+\ket{lmn}_1\ket{kl}_2\ket{kmn}_3
\big) 
+\frac{24576}{13}\delta^{ij}\ket{kl}_1\ket{kmn}_2\ket{lmn}_3
-\frac{135168}{13}\ket{ij}_1\ket{klm}_2\ket{klm}_3
\]
\[
+\frac{12288}{13}\big(
\ket{klm}_1\ket{ij}_2\ket{klm}_3+\ket{klm}_1\ket{klm}_2\ket{ij}_3
+\ket{klm}_1\ket{lmi}_2\ket{kj}_3+\ket{klm}_1\ket{lmj}_2\ket{ki}_3
+\ket{klm}_1\ket{mi}_2\ket{klj}_3+\ket{klm}_1\ket{mj}_2\ket{kli}_3\big)
\]
\[
-12288\big(\ket{kmi}_1\ket{lmj}_2\ket{kl}_3+\ket{kmj}_1\ket{lmi}_2\ket{kl}_3
+\ket{kli}_1\ket{lm}_2\ket{kmj}_3+\ket{klj}_1\ket{lm}_2\ket{kmi}_3\big)
-3072\big(\ket{kl}_1\ket{kmi}_2\ket{lmj}_3+\ket{kl}_1\ket{kmj}_2\ket{lmi}_3\big)
\]
\[
-\frac{87552}{13}
\big(\ket{kli}_1\ket{klm}_2\ket{mj}_3+\ket{klj}_1\ket{klm}_2\ket{mi}_3
+\ket{kli}_1\ket{mj}_2\ket{klm}_3+\ket{klj}_1\ket{mi}_2\ket{klm}_3\big)
\]
\beq
-\frac{35328}{13}
\big(\ket{ki}_1\ket{lmj}_2\ket{klm}_3+\ket{kj}_1\ket{lmi}_2\ket{klm}_3
+\ket{ki}_1\ket{klm}_2\ket{lmj}_3+\ket{kj}_1\ket{klm}_2\ket{lmi}_3\big),
\eqnum
\eeq
}
\pagebreak
{ \tiny
\[
\ket{\phi'_{(3)}{}_{12}^{ij}} =  
\frac{15360}{13}\big[
(\gamma^{ki})_{\alpha\beta}\ket{\alpha j}_1\ket{\beta l}_2\ket{kl}_3
+(\gamma^{kj})_{\alpha\beta}\ket{\alpha i}_1\ket{\beta l}_2\ket{kl}_3
+(\gamma^{ki})_{\alpha\beta}\ket{\alpha l}_1\ket{\beta j}_2\ket{kl}_3
+(\gamma^{kj})_{\alpha\beta}\ket{\alpha l}_1\ket{\beta i}_2\ket{kl}_3
\big]
\]
\[
-\frac{128000}{117}\big[
(\gamma^{ki})_{\alpha\beta}\ket{\alpha l}_1\ket{\beta l}_2\ket{kj}_3
+(\gamma^{kj})_{\alpha\beta}\ket{\alpha l}_1\ket{\beta l}_2\ket{ki}_3
\big]
\]
\[
+\frac{204800}{117}\big[
\ket{\alpha k}_1\ket{\alpha i}_2\ket{kj}_3+\ket{\alpha k}_1\ket{\alpha j}_2\ket{ki}_3
-\ket{\alpha i}_1\ket{\alpha k}_2\ket{kj}_3-\ket{\alpha j}_1\ket{\alpha k}_2\ket{ki}_3
\big]
\]
\[
-\frac{512}{13}\sqrt{3}\delta^{ij}(\gamma^{klm})_{\alpha\beta}
\ket{\alpha n}_1\ket{\beta n}_2\ket{klm}_3
+\frac{48128}{39 \sqrt{3}}\delta^{ij}(\gamma^{k})_{\alpha\beta}
\ket{\alpha l}_1\ket{\beta m}_2\ket{klm}_3
\]
\[
-\frac{41216}{39\sqrt{3}}\big[
(\gamma^{kli})_{\alpha\beta}\ket{\alpha j}_1\ket{\beta m}_2\ket{klm}_3
+(\gamma^{klj})_{\alpha\beta}\ket{\alpha i}_1\ket{\beta m}_2\ket{klm}_3
+(\gamma^{kli})_{\alpha\beta}\ket{\alpha m}_1\ket{\beta j}_2\ket{klm}_3
+(\gamma^{klj})_{\alpha\beta}\ket{\alpha m}_1\ket{\beta i}_2\ket{klm}_3
\big]
\]
\[
+\frac{8704}{39\sqrt{3}}(\gamma^{klm})_{\alpha\beta}\big[
\ket{\alpha i}_1\ket{\beta j}_2\ket{klm}_3
+\ket{\alpha j}_1\ket{\beta i}_2\ket{klm}_3
\big]
+\frac{12032}{39\sqrt{3}}\big[
(\gamma^{kli})_{\alpha\beta}\ket{\alpha m}_1\ket{\beta m}_2\ket{klj}_3
+(\gamma^{klj})_{\alpha\beta}\ket{\alpha m}_1\ket{\beta m}_2\ket{kli}_3
\big]
\]
\beq
-\frac{9728}{3\sqrt{3}}\big[
(\gamma^{i})_{\alpha\beta}\ket{\alpha k}_1\ket{\beta l}_2\ket{klj}_3
+(\gamma^{j})_{\alpha\beta}\ket{\alpha k}_1\ket{\beta l}_2\ket{kli}_3
\big]
+\frac{37376}{39\sqrt{3}}(\gamma^{k})_{\alpha\beta}\big[
\ket{\alpha l}_1\ket{\beta i}_2\ket{klj}_3+\ket{\alpha l}_1\ket{\beta j}_2\ket{kli}_3
-\ket{\alpha i}_1\ket{\beta l}_2\ket{klj}_3-\ket{\alpha j}_1\ket{\beta l}_2\ket{kli}_3\big],
\eqnum
\eeq
}
\bea
\ket{\phi'_{(3)}{}_{22}^{ij}}
=\ket{\phi'_{(3)}{}_{11}^{ij}}\Big|_{
\ket{*_1}_1\ket{*_2}_2\ket{*_3}_3\rightarrow \ket{*_1}_2\ket{*_2}_3\ket{*_3}_1},
\\
\ket{\phi'_{(3)}{}_{33}^{ij}}
=\ket{\phi'_{(3)}{}_{11}^{ij}}\Big|_{
\ket{*_1}_1\ket{*_2}_2\ket{*_3}_3\rightarrow \ket{*_1}_3\ket{*_2}_1\ket{*_3}_2},
\\
\ket{\phi'_{(3)}{}_{23}^{ij}}
=\ket{\phi'_{(3)}{}_{12}^{ij}}\Big|_{
\ket{*_1}_1\ket{*_2}_2\ket{*_3}_3\rightarrow \ket{*_1}_2\ket{*_2}_3\ket{*_3}_1},
\\
\ket{\phi'_{(3)}{}_{31}^{ij}}
=\ket{\phi'_{(3)}{}_{12}^{ij}}\Big|_{
\ket{*_1}_1\ket{*_2}_2\ket{*_3}_3\rightarrow \ket{*_1}_3\ket{*_2}_1\ket{*_3}_2}.
\eea
Unlike $\ket{\phi_{(2)}{}_{11}^{ij}}$, $\ket{\phi'_{(3)}{}_{11}^{ij}}$
does not have terms in the form of $\ket{k_1k_2k_3}_1\ket{k_4k_5k_6}_2\ket{k_7k_8k_9}_3$.
This shows that $\ket{\phi_{(2)}{}_{ab}^{ij}}$ and $\ket{\phi_{(3)}{}_{ab}^{ij}}$ are independent of each other.

Noting that $\theta^a\gamma^{i}{}_{n_1}\theta^a$ is an $SO(9)$ generator, we obtain
\bea
\ket{\phi'_{(3)}{}^{ij}_{aa}} & = &
72[\theta^c\gamma^{(i}{}_{n_2n_3}\theta^c][\theta^d\gamma^{j)n_2n_3}\theta^d]\ket{S}
-8\delta^{ij}[\theta^c\gamma_{n_1n_2n_3}\theta^c][\theta^d\gamma^{n_1n_2n_3}\theta^d]\ket{S}.
\eea
This shows that $\ket{\phi'_{(3)}{}^{ii}_{aa}}=0$, and 
therefore $\ket{\phi'_{(3)}{}^{ii}_{ab}}$ is in $(s,t)$ representation. 
However there is no $(s,t)$ representation in the table in \cite{construction3}. 
This means $\ket{\phi'_{(3)}{}^{ii}_{ab}}=0$.
The above explicit expression of $\ket{\phi'_{(3)}{}^{ij}_{ab}}$ indeed satisfies this, and 
the traceless part $\ket{\phi_{(3)}{}^{ij}_{ab}}$ is defined by
\beq
\ket{\phi_{(3)}{}^{ij}_{ab}}:=\ket{\phi'_{(3)}{}^{ij}_{ab}}-\frac{1}{3}\delta^{ab}\ket{\phi'_{(3)}{}^{ij}_{cc}}.
\eeq
Let us make another check on the above expression of $\ket{\phi'_{(3)}{}^{ij}_{ab}}$:
$\ket{\phi'_{(3)}{}^{ij}_{aa}}$ gives a $(t,s)$ representation, and since this representation must be unique,
this must be proportional to $\ket{\phi{}^{ij}}$. Indeed,
\beq
\ket{\phi'_{(3)}{}^{ij}_{aa}}=-288\ket{\phi{}^{ij}}.
\eeq
Explicit expression of $\ket{\phi_{(3)}{}^{ij}_{ab}}$ is given in the Appendix.

%%%%%%%%%%%%%%%%%%%%%%%%%%%%%%%%%%%%%%%%%%%%%%%%%%%%%%%%%%%%%%%%%%%
\subsection{Solutions to Schr\"odinger equation}

Next we construct solutions to zero-energy Schr\"odinger equation 
$(\gamma^i\theta^a)_\alpha\ket{\phi_{ab}^{ij}}=0$.
The left hand side of this equation has one vector index and one spinor index of $SO(9)$, and one adjoint index of $SU(2)$.
This $SO(9)$ representation can be decomposed into a vector-spinor($\bold{128}$) and a spinor($\bold{16}$) representation.
According to the table in \cite{construction3} there is a unique spinor representation, and there are two independent vector-spinor
representations. Therefore this equation can be decomposed into three equations for those three representations.
$\ket{\phi_{ab}^{ij}}$ is given as a linear combination of five states we have constructed:
\beq
\ket{\phi_{ab}^{ij}}=c_1\delta_{ab}\delta^{ij}\ket{S}+c_2\delta_{ab}\ket{\phi^{ij}}
+c_3\ket{\phi_{(1)}{}_{ab}^{ij}}+c_4\ket{\phi_{(2)}{}_{ab}^{ij}}+c_5\ket{\phi_{(3)}{}_{ab}^{ij}},
\eeq
and three of those five coefficients $c_i$ can be determined. This means that we have two independent solutions to
the Schr\"odinger equation.

First we show $c_1=0$ with a shorter calculation:
Since $(\theta^b\gamma^j\gamma^i\theta^a)\ket{\phi_{ab}^{ij}}=
\theta^b_\alpha\theta^a_\alpha\ket{\phi_{ab}^{ii}}+(\theta^b\gamma^{ji}\theta^a)_\alpha\ket{\phi_{ab}^{ij}}
=\theta^b_\alpha\theta^a_\alpha\ket{\phi_{ab}^{ii}}
+(\theta^{(b}\gamma^{ji}\theta^{a)})_\alpha\ket{\phi_{ab}^{ij}}$,
four of five states
$\ket{\phi^{ij}}$, $\ket{\phi_{(1)}{}_{ab}^{ij}}$, $\ket{\phi_{(2)}{}_{ab}^{ij}}$, and $\ket{\phi_{(3)}{}_{ab}^{ij}}$ do not contribute to
$(\theta^b\gamma^j\gamma^i\theta^a)\ket{\phi_{ab}^{ij}}$, as can be easily seen from the symmetry of indices.
Therefore $(\theta^b\gamma^j\gamma^i\theta^a)\ket{\phi_{ab}^{ij}}=0$ gives $c_1=0$.

Then we deal with $(\gamma^i\theta^a)_\alpha\ket{\phi_{ab}^{ij}}=0$ directly.
The coefficients of independent states in explicit expression of $(\gamma^i\theta^a)_\alpha\ket{\phi_{ab}^{ij}}$
must give only two independent equations for $c_i$. There are more than 40 independent states in
$(\gamma^i\theta^a)_\alpha\ket{\phi_{ab}^{ij}}$, and we have checked that all of them reduce to
the following two equations:
\beq
c_2=\frac{7}{20}c_4 + 240c_5,\quad c_3=\frac{31}{30}c_4 - 96c_5.
\eeq
Then we obtain two independent solutions:
\bea
\frac{7}{20}\delta_{ab}\ket{\phi^{ij}}
+\frac{31}{30}\ket{\phi_{(1)}{}_{ab}^{ij}}+\ket{\phi_{(2)}{}_{ab}^{ij}},
\\
240\delta_{ab}\ket{\phi^{ij}}
-96\ket{\phi_{(1)}{}_{ab}^{ij}}+\ket{\phi_{(3)}{}_{ab}^{ij}}.
\eea

%%%%%%%%%%%%%%%%%%%%%%%%%%%%%%%%%%%%%%%%%%%%%%%%%%%%%%%%%%%%%%%%%%%
%%%%%%%%%%%%%%%%%%%%%%%%%%%%%%%%%%%%%%%%%%%%%%%%%%%%%%%%%%%%%%%%%%%
\section{Discussion}

Zero-energy states of supersymmetric models can often be found explicitly due to the fact that they satisfy first order differential equations. 
It is clear however that for $\mathcal{N}=16$ supermembrane matrix model this simplification is not enough to find the corresponding wave-function - it seems not likely that one can just guess the form of the state. For this reason the Taylor expansion approach initiated in \cite{construction1} is a natural step forward. The 2nd order terms, determined in this paper, together with the 0th \cite{construction1} and the 1st \cite{construction4} order ones complete the initial conditions needed to solve the recurrence relation (\ref{rest}) for all higher terms. It is therefore a crucial step towards finding the ground state by this method.

We conjecture that the ground state can be written in terms of variables $X$ and $\theta$ acting on $\ket{S}$ as in (\ref{ansatz}). This statement is true for the 0th, 1st and the 2nd order terms and since they provide the initial conditions for the recurrence relation (\ref{rest}) it is very likely that it holds for all other terms.

Although the numbers of representations necessary for constructing
solutions to zero energy Schr\"odinger equation increase as we go to
higher order, we have found that it is not so time-consuming to solve
the equation at each order with present PC's in the case of $SU(2)$
gauge group.

%%%%%%%%%%%%%%%%%%%%%%%%%%%%%%%%%%%%%%%%%%%%%%%%%%%%%%%%%%%%%%%
\vs{.5cm}
\noindent
{\large\bf Acknowledgements}\\[.2cm]
We would like to thank J.\ Hoppe, M.\ Hynek and D.\ Lundholm for discussions, and 
Y. M. would like to thank Max Planck Institute for Gravitational Physics
(Albert Einstein Institute) for kind hospitality.
This work was supported by DFG (German Science Foundation) via the SFB grant (MT).
%%%%%%%%%%%%%%%%%%%%%%%%%%%%%%%%%%%%%%%%%%%%%%%%%%%%%%%%%%%%%%%

\renewcommand{\theequation}{\Alph{section}.\arabic{equation}}
\appendix
\addcontentsline{toc}{section}{Appendix}

\renewcommand{\thesection}{Appendix:}
%%%%%%%%%%%%%%%%%%%%%%%%%%%%%%%%%%%%%%%%%%%%%%%%%%%%%%%%%%%%%
\vs{.5cm}
\noindent
\section{Summary of Results}
\label{appa}
\setcounter{equation}{0}
In this appendix we give a summary of explicit expressions of 
$\ket{\phi^{ij}}$, $\ket{\phi_{(1)}{}_{ab}^{ij}}$, $\ket{\phi_{(2)}{}_{ab}^{ij}}$, and
$\ket{\phi_{(3)}{}_{ab}^{ij}}$.
\bea
\ket{\phi^{ij}} & := & \big[O_c^{(i}O_c^{j)}+160\delta^{ij}\big]\ket{S}
\nn & = &
\frac{1}{13}\Big[
-992\big(
\ket{ij}_1\ket{kl}_2\ket{kl}_3 +\ket{kl}_1\ket{ij}_2\ket{kl}_3 +\ket{kl}_1\ket{kl}_2\ket{ij}_3 
\big) \nn & &
+96\big(
\ket{ki}_1\ket{lj}_2\ket{kl}_3 +\ket{lj}_1\ket{kl}_2\ket{ki}_3 +\ket{kl}_1\ket{ki}_2\ket{lj}_3 
 \nn & &
+\ket{kj}_1\ket{li}_2\ket{kl}_3 +\ket{li}_1\ket{kl}_2\ket{kj}_3 +\ket{kl}_1\ket{kj}_2\ket{li}_3 
\big) \nn & &
-64\delta^{ij}\ket{kl}_1\ket{lm}_2\ket{mk}_3
 \nn & &
-448\delta^{ij}\big(
\ket{kmn}_1\ket{lmn}_2\ket{kl}_3 +\ket{lmn}_1\ket{kl}_2\ket{kmn}_3 +\ket{kl}_1\ket{kmn}_2\ket{lmn}_3 
\big)\nn & &
+384\big(
\ket{klm}_1\ket{klm}_2\ket{ij}_3 
+\ket{klm}_1\ket{ij}_2\ket{klm}_3 
+\ket{ij}_1\ket{klm}_2\ket{klm}_3 
 \nn & &
+\ket{ki}_1\ket{klm}_2\ket{lmj}_3
+\ket{klj}_1\ket{mi}_2\ket{klm}_3 
+\ket{klm}_1\ket{lmj}_2\ket{ki}_3 
 \nn & &
+\ket{kj}_1\ket{klm}_2\ket{lmi}_3 
+\ket{kli}_1\ket{mj}_2\ket{klm}_3 
+\ket{klm}_1\ket{lmi}_2\ket{kj}_3 
 \nn & &
+\ket{ki}_1\ket{lmj}_2\ket{klm}_3 
+\ket{klm}_1\ket{mi}_2\ket{klj}_3 
+\ket{klj}_1\ket{klm}_2\ket{mi}_3 
 \nn & &
+\ket{kj}_1\ket{lmi}_2\ket{klm}_3 
+\ket{klm}_1\ket{mj}_2\ket{kli}_3 
+\ket{kli}_1\ket{klm}_2\ket{mj}_3 
\big) \nn & &
+1248\big(
\ket{kmi}_1\ket{lmj}_2\ket{kl}_3 +\ket{lmj}_1\ket{kl}_2\ket{kmi}_3 +\ket{kl}_1\ket{kmi}_2\ket{lmj}_3 
\nn & &
+\ket{kmj}_1\ket{lmi}_2\ket{kl}_3 +\ket{lmi}_1\ket{kl}_2\ket{kmj}_3 +\ket{kl}_1\ket{kmj}_2\ket{lmi}_3 
\big)\Big].
\eea

\beq
\ket{\phi_{(1)}{}_{ab}^{ij}} := O_{[a}^{[i}O_{b]}^{j]}\ket{S}.
\eeq
\bea
\ket{\phi_{(1)}{}_{12}^{ij}} & = & 
-\frac{160}{9}(\gamma^{k[i})_{\alpha\beta}\big(
\ket{\alpha j]}_1\ket{\beta l}_2\ket{kl}_3-\ket{\alpha l}_1\ket{\beta j]}_2\ket{kl}_3
\big) \nn & & +
\sqrt{3}\Big[\frac{112}{9}(\gamma^{kl[i})_{\alpha\beta}\big(
\ket{\alpha j]}_1\ket{\beta m}_2\ket{klm}_3-\ket{\alpha m}_1\ket{\beta j]}_2\ket{klm}_3
\big) \nn & & 
-\frac{32}{27}(\gamma^{k})_{\alpha\beta}\big(
\ket{\alpha [i}_1\ket{\beta l}_2\ket{klj]}_3+\ket{\alpha l}_1\ket{\beta [i}_2\ket{klj]}_3
\big) \nn & & 
+\frac{304}{27}(\gamma^{kij})_{\alpha\beta}\ket{\alpha l}_1\ket{\beta m}_2\ket{klm}_3
 \nn & & 
-\frac{40}{81}(\gamma^{klmij})_{\alpha\beta}\ket{\alpha n}_1\ket{\beta n}_2\ket{klm}_3
 \nn & & 
+\frac{272}{81}(\gamma^{klm})_{\alpha\beta}\ket{\alpha [i}_1\ket{\beta j]}_2\ket{klm}_3
 \nn & & 
+\frac{416}{27}(\gamma^{k})_{\alpha\beta}\ket{\alpha l}_1\ket{\beta l}_2\ket{kij}_3
\Big],
\\
\ket{\phi_{(1)}{}_{23}^{ij}} & = & \ket{\phi_{(1)}{}_{12}^{ij}}\Big|_{
\ket{*_1}_1\ket{*_2}_2\ket{*_3}_3\rightarrow \ket{*_1}_2\ket{*_2}_3\ket{*_3}_1},
\\
\ket{\phi_{(1)}{}_{31}^{ij}} & = & \ket{\phi_{(1)}{}_{12}^{ij}}\Big|_{
\ket{*_1}_1\ket{*_2}_2\ket{*_3}_3\rightarrow \ket{*_1}_3\ket{*_2}_1\ket{*_3}_2}.
\eea
\beq
\ket{\phi_{(2)}{}_{ab}^{ij}} := \big[O_{(a}^{(i}O_{b)}^{j)}-\frac{1}{3}\delta^{ab}O_{c}^{(i}O_{c}^{j)}\big]\ket{S}.
\eeq
\bea
\ket{\phi_{(2)}{}_{11}^{ij}} & = &
\frac{32}{13}\big(\ket{kl}_1\ket{kl}_2\ket{ij}_3+\ket{kl}_1\ket{ij}_2\ket{kl}_3
-2\ket{ij}_1\ket{kl}_2\ket{kl}_3 \big)
\nn & &
+\frac{944}{39}\big(\ket{ki}_1\ket{lj}_2\ket{kl}_3+\ket{kj}_1\ket{kl}_2\ket{li}_3
-2\ket{kl}_1\ket{li}_2\ket{kj}_3
\nn & &
+\ket{kj}_1\ket{li}_2\ket{kl}_3+\ket{ki}_1\ket{kl}_2\ket{lj}_3
-2\ket{kl}_1\ket{lj}_2\ket{ki}_3
\big)
\nn & &
-\frac{592}{39}\delta^{ij}\big(\ket{klm}_1\ket{lmn}_2\ket{kn}_3+\ket{klm}_1\ket{mn}_2\ket{kln}_3
-2\ket{kl}_1\ket{lmn}_2\ket{kmn}_3\big)
\nn & &
-\frac{128}{13}\big(\ket{klm}_1\ket{klm}_2\ket{ij}_3+\ket{klm}_1\ket{ij}_2\ket{klm}_3
 -2\ket{ij}_1\ket{klm}_2\ket{klm}_3 \big)
\nn & &
+16\big(\ket{kli}_1\ket{kmj}_2\ket{lm}_3+\ket{klj}_1\ket{lm}_2\ket{kmi}_3
-2\ket{kl}_1\ket{lmi}_2\ket{kmj}_3
\nn & &
+\ket{klj}_1\ket{kmi}_2\ket{lm}_3+\ket{kli}_1\ket{lm}_2\ket{kmj}_3
-2\ket{kl}_1\ket{lmj}_2\ket{kmi}_3
\big)
\nn & &
+\frac{704}{13}\big(
\ket{klm}_1\ket{lmi}_2\ket{kj}_3+\ket{klm}_1\ket{lmj}_2\ket{ki}_3
\nn & &
+\ket{klm}_1\ket{mi}_2\ket{klj}_3+\ket{klm}_1\ket{mj}_2\ket{kli}_3
\big)
\nn & &
-\frac{24}{13}\big(
\ket{kli}_1\ket{klm}_2\ket{mj}_3+\ket{klj}_1\ket{klm}_2\ket{mi}_3
\nn & &
+\ket{kli}_1\ket{mj}_2\ket{klm}_3+\ket{klj}_1\ket{mi}_2\ket{klm}_3
\big)
\nn & &
-\frac{680}{13}\big(
\ket{ki}_1\ket{lmj}_2\ket{klm}_3+\ket{kj}_1\ket{lmi}_2\ket{klm}_3
\nn & &
+\ket{ki}_1\ket{klm}_2\ket{lmj}_3+\ket{kj}_1\ket{klm}_2\ket{lmi}_3
\big)
\nn & &
-\frac{8}{3\sqrt{3}}\big(
\epsilon^{k_1k_2k_3k_4k_5k_6k_7k_8i}\ket{k_1k_2k_3}_1\ket{k_4k_5j}_2\ket{k_6k_7k_8}_3
\nn & &
+\epsilon^{k_1k_2k_3k_4k_5k_6k_7k_8j}\ket{k_1k_2k_3}_1\ket{k_4k_5i}_2\ket{k_6k_7k_8}_3
\nn & &
+\epsilon^{k_1k_2k_3k_4k_5k_6k_7k_8i}\ket{k_1k_2k_3}_1\ket{k_4k_5k_6}_2\ket{k_7k_8j}_3
\nn & &
+\epsilon^{k_1k_2k_3k_4k_5k_6k_7k_8j}\ket{k_1k_2k_3}_1\ket{k_4k_5k_6}_2\ket{k_7k_8i}_3
\big),
\\
\ket{\phi_{(2)}{}_{12}^{ij}}  & = & 
-\frac{1}{351}\Big[
288(\gamma^{k(i})_{\alpha\beta}\big(
\ket{\alpha j)}_1\ket{\beta l}_2\ket{kl}_3+\ket{\alpha l}_1\ket{\beta j)}_2\ket{kl}_3
\big) \nn & &
+4448(\gamma^{k(i})_{\alpha\beta}\ket{\alpha l}_1\ket{\beta l}_2\ket{kj)}_3 \nn & &
+10112\big(
\ket{\alpha (i}_1\ket{\alpha k}_2\ket{kj)}_3-\ket{\alpha k}_1\ket{\alpha (i}_2\ket{j)k}_3
\big)
\Big] \nn & &
+\frac{\sqrt{3}}{351}\Big[
904(\gamma^{kl(i})_{\alpha\beta}\big(
\ket{\alpha j)}_1\ket{\beta m}_2\ket{klm}_3+\ket{\alpha m}_1\ket{\beta j)}_2\ket{klm}_3
\big) \nn & &
-5552(\gamma^{k})_{\alpha\beta}\big(
\ket{\alpha (i}_1\ket{\beta l}_2\ket{klj)}_3-\ket{\alpha l}_1\ket{\beta (i}_2\ket{klj)}_3
\big) \nn & &
-1520\delta^{ij}(\gamma^{k})_{\alpha\beta}\ket{\alpha l}_1\ket{\beta m}_2\ket{klm}_3 \nn & &
-120\delta^{ij}(\gamma^{klm})_{\alpha\beta}\ket{\alpha n}_1\ket{\beta n}_2\ket{klm}_3 \nn & &
-656(\gamma^{klm})_{\alpha\beta}\ket{\alpha (i}_1\ket{\beta j)}_2\ket{klm}_3 \nn & &
-1040(\gamma^{(i})_{\alpha\beta}\ket{\alpha k}_1\ket{\beta l}_2\ket{klj)}_3 \nn & &
+1736(\gamma^{kl(i})_{\alpha\beta}\ket{\alpha n}_1\ket{\beta n}_2\ket{klj)}_3
\Big],
\\
\ket{\phi_{(2)}{}_{22}^{ij}} & = & \ket{\phi_{(2)}{}_{11}^{ij}}\Big|_{
\ket{*_1}_1\ket{*_2}_2\ket{*_3}_3\rightarrow \ket{*_1}_2\ket{*_2}_3\ket{*_3}_1},
\\
\ket{\phi_{(2)}{}_{33}^{ij}} & = & \ket{\phi_{(2)}{}_{11}^{ij}}\Big|_{
\ket{*_1}_1\ket{*_2}_2\ket{*_3}_3\rightarrow \ket{*_1}_3\ket{*_2}_1\ket{*_3}_2},
\\
\ket{\phi_{(2)}{}_{23}^{ij}} & = & \ket{\phi_{(2)}{}_{12}^{ij}}\Big|_{
\ket{*_1}_1\ket{*_2}_2\ket{*_3}_3\rightarrow \ket{*_1}_2\ket{*_2}_3\ket{*_3}_1},
\\
\ket{\phi_{(2)}{}_{31}^{ij}} & = & \ket{\phi_{(2)}{}_{12}^{ij}}\Big|_{
\ket{*_1}_1\ket{*_2}_2\ket{*_3}_3\rightarrow \ket{*_1}_3\ket{*_2}_1\ket{*_3}_2}.
\eea

\beq
\ket{\phi_{(3)}{}_{ab}^{ij}} :=
 \ket{\phi'_{(3)}{}_{ab}^{ij}}-\frac{1}{3}\delta^{ab}\ket{\phi'_{(3)}{}_{cc}^{ij}}.
\eeq
\bea
\ket{\phi_{(3)}{}_{11}^{ij}} & = &
-\frac{92160}{13}\big(\ket{kl}_1\ket{ij}_2\ket{kl}_3+\ket{kl}_1\ket{kl}_2\ket{ij}_3
-2\ket{ij}_1\ket{kl}_2\ket{kl}_3\big)
\nn & &
+\frac{25600}{13}\big(
\ket{ki}_1\ket{lj}_2\ket{kl}_3
+\ket{kj}_1\ket{kl}_2\ket{li}_3-2\ket{kl}_1\ket{li}_2\ket{kj}_3
\nn & &
+\ket{kj}_1\ket{li}_2\ket{kl}_3+\ket{ki}_1\ket{kl}_2\ket{lj}_3
-2\ket{kl}_1\ket{lj}_2\ket{ki}_3
\big)
\nn & &
+\frac{9216}{13}\delta^{ij}\big(\ket{klm}_1\ket{lmn}_2\ket{kn}_3+\ket{klm}_1\ket{mn}_2\ket{kln}_3
 -2\ket{kl}_1\ket{lmn}_2\ket{kmn}_3\big)
\nn & &
+3072\big(
\ket{kli}_1\ket{lmj}_2\ket{km}_3+\ket{klj}_1\ket{km}_2\ket{lmi}_3-2\ket{kl}_1\ket{lmi}_2\ket{kmj}_3
\nn & &
+\ket{klj}_1\ket{lmi}_2\ket{km}_3+\ket{kli}_1\ket{km}_2\ket{lmj}_3-2\ket{kl}_1\ket{lmj}_2\ket{kmi}_3
\big)
\nn & &
+\frac{49152}{13}\big(
\ket{klm}_1\ket{klm}_2\ket{ij}_3
+\ket{klm}_1\ket{ij}_2\ket{klm}_3-2\ket{ij}_1\ket{klm}_2\ket{klm}_3
\nn & &
+\ket{klm}_1\ket{lmi}_2\ket{kj}_3+\ket{klm}_1\ket{lmj}_2\ket{ki}_3
\nn & &
+\ket{klm}_1\ket{mi}_2\ket{klj}_3+\ket{klm}_1\ket{mj}_2\ket{kli}_3
\big)
\nn & &
-\frac{50688}{13}\big(
\ket{kli}_1\ket{klm}_2\ket{mj}_3
+\ket{klj}_1\ket{klm}_2\ket{mi}_3
\nn & &
+\ket{kli}_1\ket{mj}_2\ket{klm}_3
+\ket{klj}_1\ket{mi}_2\ket{klm}_3
\big)
\nn & &
+\frac{1536}{13}\big(
\ket{ki}_1\ket{lmj}_2\ket{klm}_3
+\ket{kj}_1\ket{lmi}_2\ket{klm}_3
\nn & &
+\ket{ki}_1\ket{klm}_2\ket{lmj}_3
+\ket{kj}_1\ket{klm}_2\ket{lmi}_3
\big),
\\
\ket{\phi_{(3)}{}_{12}^{ij}} & = & 
\frac{15360}{13}\big[
(\gamma^{ki})_{\alpha\beta}\ket{\alpha j}_1\ket{\beta l}_2\ket{kl}_3
+(\gamma^{kj})_{\alpha\beta}\ket{\alpha i}_1\ket{\beta l}_2\ket{kl}_3
\nn & &
+(\gamma^{ki})_{\alpha\beta}\ket{\alpha l}_1\ket{\beta j}_2\ket{kl}_3
+(\gamma^{kj})_{\alpha\beta}\ket{\alpha l}_1\ket{\beta i}_2\ket{kl}_3
\big]
\nn & &
-\frac{128000}{117}\big[
(\gamma^{ki})_{\alpha\beta}\ket{\alpha l}_1\ket{\beta l}_2\ket{kj}_3
+(\gamma^{kj})_{\alpha\beta}\ket{\alpha l}_1\ket{\beta l}_2\ket{ki}_3
\big]
\nn & &
+\frac{204800}{117}\big[
\ket{\alpha k}_1\ket{\alpha i}_2\ket{kj}_3+\ket{\alpha k}_1\ket{\alpha j}_2\ket{ki}_3
\nn & &
-\ket{\alpha i}_1\ket{\alpha k}_2\ket{kj}_3-\ket{\alpha j}_1\ket{\alpha k}_2\ket{ki}_3
\big]
\nn & &
-\frac{512}{13}\sqrt{3}\delta^{ij}(\gamma^{klm})_{\alpha\beta}
\ket{\alpha n}_1\ket{\beta n}_2\ket{klm}_3
\nn & &
+\frac{48128}{39 \sqrt{3}}\delta^{ij}(\gamma^{k})_{\alpha\beta}
\ket{\alpha l}_1\ket{\beta m}_2\ket{klm}_3
\nn & &
-\frac{41216}{39\sqrt{3}}\big[
(\gamma^{kli})_{\alpha\beta}\ket{\alpha j}_1\ket{\beta m}_2\ket{klm}_3
+(\gamma^{klj})_{\alpha\beta}\ket{\alpha i}_1\ket{\beta m}_2\ket{klm}_3
\nn & &
+(\gamma^{kli})_{\alpha\beta}\ket{\alpha m}_1\ket{\beta j}_2\ket{klm}_3
+(\gamma^{klj})_{\alpha\beta}\ket{\alpha m}_1\ket{\beta i}_2\ket{klm}_3
\big]
\nn & &
+\frac{8704}{39\sqrt{3}}(\gamma^{klm})_{\alpha\beta}\big[
\ket{\alpha i}_1\ket{\beta j}_2\ket{klm}_3
+\ket{\alpha j}_1\ket{\beta i}_2\ket{klm}_3
\big]
\nn & &
+\frac{12032}{39\sqrt{3}}\big[
(\gamma^{kli})_{\alpha\beta}\ket{\alpha m}_1\ket{\beta m}_2\ket{klj}_3
+(\gamma^{klj})_{\alpha\beta}\ket{\alpha m}_1\ket{\beta m}_2\ket{kli}_3
\big]
\nn & &
-\frac{9728}{3\sqrt{3}}\big[
(\gamma^{i})_{\alpha\beta}\ket{\alpha k}_1\ket{\beta l}_2\ket{klj}_3
+(\gamma^{j})_{\alpha\beta}\ket{\alpha k}_1\ket{\beta l}_2\ket{kli}_3
\big]
\nn & &
+\frac{37376}{39\sqrt{3}}(\gamma^{k})_{\alpha\beta}\big[
\ket{\alpha l}_1\ket{\beta i}_2\ket{klj}_3+\ket{\alpha l}_1\ket{\beta j}_2\ket{kli}_3
\nn & &
-\ket{\alpha i}_1\ket{\beta l}_2\ket{klj}_3-\ket{\alpha j}_1\ket{\beta l}_2\ket{kli}_3\big],
\\
\ket{\phi_{(3)}{}_{22}^{ij}} & = & \ket{\phi_{(3)}{}_{11}^{ij}}\Big|_{
\ket{*_1}_1\ket{*_2}_2\ket{*_3}_3\rightarrow \ket{*_1}_2\ket{*_2}_3\ket{*_3}_1},
\\
\ket{\phi_{(3)}{}_{33}^{ij}} & = & \ket{\phi_{(3)}{}_{11}^{ij}}\Big|_{
\ket{*_1}_1\ket{*_2}_2\ket{*_3}_3\rightarrow \ket{*_1}_3\ket{*_2}_1\ket{*_3}_2},
\\
\ket{\phi_{(3)}{}_{23}^{ij}} & = & \ket{\phi_{(3)}{}_{12}^{ij}}\Big|_{
\ket{*_1}_1\ket{*_2}_2\ket{*_3}_3\rightarrow \ket{*_1}_2\ket{*_2}_3\ket{*_3}_1},
\\
\ket{\phi_{(3)}{}_{31}^{ij}} & = & \ket{\phi_{(3)}{}_{12}^{ij}}\Big|_{
\ket{*_1}_1\ket{*_2}_2\ket{*_3}_3\rightarrow \ket{*_1}_3\ket{*_2}_1\ket{*_3}_2}.
\eea

%%%%%%%%%%%% References %%%%%%%%%%%%%%%%%%%%%%%%%
\newcommand{\J}[4]{{\sl #1} {\bf #2} (#3) #4}
\newcommand{\andJ}[3]{{\bf #1} (#2) #3}
\newcommand{\AP}{Ann.\ Phys.\ (N.Y.)}
\newcommand{\MPL}{Mod.\ Phys.\ Lett.}
\newcommand{\NP}{Nucl.\ Phys.}
\newcommand{\PL}{Phys.\ Lett.}
\newcommand{\PR}{Phys.\ Rev.}
\newcommand{\PRL}{Phys.\ Rev.\ Lett.}
\newcommand{\PTP}{Prog.\ Theor.\ Phys.}
\newcommand{\hepth}[1]{{\tt hep-th/#1}}
\newcommand{\arxivhep}[1]{{\tt arXiv.org:#1 [hep-th]}}
%%%%%%%%%%%%%%%%%%%%%%%%%%%%%%%%%%%%%%%%%%%%%%%%

%%%%%%%%%%%%%%%%%%%%%%%%%%%%%%%%%%%%%%%%%%%%%%%%%%%%%%%%%%%%%%%%%%%
\end{document}